\documentclass[conference]{IEEEtran}
\usepackage[pdftex]{graphicx}
\usepackage{algpseudocode}
\usepackage{amsthm}
\usepackage[sort,compress]{cite}
\usepackage{hyperref}
\hypersetup{bookmarksdepth=-2}
\usepackage{epsfig}
\usepackage{array}
\newcolumntype{L}[1]{>{\raggedright\let\newline\\\arraybackslash\hspace{0pt}}m{#1}}

\theoremstyle{definition}

\begin{document}
\title{AutoCogniSys: IoT Assisted Context-Aware Automatic Cognitive Health Assessment}

\author{\IEEEauthorblockN{Mohammad Arif Ul Alam\textsuperscript{1},
Nirmalya Roy\textsuperscript{1}, Sarah Holmes\textsuperscript{2}, Aryya Gangopadhyay\textsuperscript{1}, Elizabeth Galik\textsuperscript{3}}
\IEEEauthorblockA{\textsuperscript{1}Department of Information Systems, University of Maryland, Baltimore County\\
\textsuperscript{2}Department of Gerontology, University of Maryland, Baltimore and Baltimore County\\
\textsuperscript{3}School of Nursing, University of Maryland, Baltimore\\
alam4@umbc.edu, nroy@umbc.edu, sarah65@umbc.edu, gangopad@umbc.edu, galik@son.umaryland.edu
}
}
\maketitle

\begin{abstract}
Cognitive impairment has become epidemic in older adult population. The recent advent of tiny wearable and ambient devices, a.k.a Internet of Things (IoT) provides ample platforms for continuous functional and cognitive health assessment of older adults. In this paper, we design, implement and evaluate \emph{AutoCogniSys}, a context-aware automated cognitive health assessment system, combining the sensing powers of wearable physiological (Electrodermal Activity, Photoplethysmography) and physical (Accelerometer, Object) sensors in conjunction with ambient sensors. We design appropriate signal processing and machine learning techniques, and develop an automatic cognitive health assessment system in a natural older adults living environment. We validate our approaches using two datasets: (i) a naturalistic sensor data streams related to Activities of Daily Living and mental arousal of 22 older adults recruited in a retirement community center, individually living in their own apartments using a customized inexpensive IoT system (IRB \#HP-00064387) and (ii) a publicly available dataset for emotion detection. The performance of \emph{AutoCogniSys} attests max. 93\% of accuracy in assessing cognitive health of older adults.
\end{abstract}


\maketitle
\section{Introduction}
Cognitive deficit of older adults is one of the biggest global public health challenges in elderly care. Approximately 5.2 million people of 65 and older are suffered with any form of cognitive impairments in United States in 2012 \cite{stat12}. Dementia is one of the major causes of the cognitive impairments which is more acute among 85 and older population (50\%) \cite{stat12}. However, the costs (financial and time) of health care and long-term care for individuals with Alzheimer's (special form of dementia) or other dementias are substantial. For example, during 2016, about 15.9 million family and friends in United States provided 18.2 billion hours of unpaid assistance to those with cognitive impairments which is a contribution to the nation valued at \$230.1 billion. One the other hand, total payments for all individuals with all form of cognitive impairments are estimated at \$259 billion. Total annual payments for health care, long-term care and hospice care for people with Alzheimer's or other dementias are projected to increase from \$259 billion in 2017 to more than \$1.1 trillion in 2050. Among the above costs, a significant amount are relevant to clinical and diagnostic tests \cite{stat17}. Although clinical and diagnostic tests have become more precise in identifying dementia, studies have shown that there is a high degree of underrecognition especially in early detection. However, there are many advantages to obtaining an early and accurate diagnosis when cognitive symptoms are first noticed as the root cause findings of impairment always lessen the progress of impairment status and sometimes symptoms can be reversible and cured.

With the proliferation of emerging ubiquitous computing technologies, many mobile and wearable devices have been available to capture continuous functional and physiological behavior of older adults. Wearable sensors are now capable of estimating number of steps being taken, physical activity levels, sleep patterns and physiological outcomes (heart rate, skin conductance) of older adults \cite{sano15}. Ambient sensors also help capture the movement patterns of objects and humans for activity and behavior recognition \cite{dawadi14,dawadi15}. Researchers also proved the existence of correlations between cognitive impairment and everyday task performance \cite{dawadi14, akl15,alam16} as well as physiological symptoms \cite{alam16,sano15}. Although current studies showed some successes in IoT-assisted cognitive health assessment in different domains individually, there are several existing challenges in developing and validating a fully automated multi-modal assessment model.

\begin{enumerate}
\item \emph{Real-time IoT System}: A real-time IoT system must include a continuous and fault tolerant data streaming capability among central hub, wearable sensors and ambient sensors regardless of network communication protocol (WiFi, Ethernet, Bluetooth etc.) which are not available in existing researches.
\item \emph{Multi-modal Context Fusion}: Though several offline clinically validated cognitive health assessment tools exist \cite{wai03, starling99, krapp07, yesavage82, zung71}, there is no universally accepted method for IoT-assisted automatic cognitive health assessment in smart home environment that can fuse multi-modal sensor contexts altogether. For example, some researchers showed ambient sensors based Activities of Daily Livigin (ADLs) sequence pattern can signify the cognitive health status of older adults \cite{akl15, dawadi15}. Researchers also showed wearable Electrodermal Activity pattern analysis may carry the significance of cognitive status \cite{sano15}. However, for validation of IoT based cognitive health assessment, self-reported surveys, clinical diagnosis and observation based tools are used individually by prior researchers \cite{akl15, dawadi15, sano15, alam16}.
\end{enumerate}

Regarding aforementioned challenges for the automation of cognitive health assessment, \emph{AutoCogniSys} considers (i) reproducibility of our model in any smart home system consists of ambient motion sensors, wearable accelerometer (ACC) sensors, wearable Electrodermal Activity (EDA) and Photoplethysmography (PPG) sensors individually or combined streams; (ii) context awareness based on ambient motion sensors and wearable ACC sensors in any types of activities such as hand gestural, postural and complex ADLs; and (iii) high accuracy, i.e., a recall rate of over 90\% with less than 5\% false positive rate. More specifically, \emph{AutoCogniSys} extends our existing work \cite{alam16} in three dimensions,

\emph{(1) True Automation:} We first investigate the correlations of cognitive impairment with human activities and stress where we manually labeled activities, extract the corresponding physiological sensor (EDA and PPG) features of each activity, and use statistical method to find correlations. Then, we propose automatic complex activity recognition based on a Hierarchical Dynamic Bayesian Network (HDBN) model, fine-grained extraction of physiological sensor features and finally machine learning classification of cognitive impairment.

\emph{(2) Noises Elimination:} We define different types of noises on ACC, EDA and PPG sensors, propose extensive signal processing techniques to remove noises and show significant improvement can be achieved in cognitive impairment classification.

\emph{(3) Implementation and Evaluation:} Finally, we design and implement IoT system and analytic methods and minimize the human involvement to automate our proposed cognitive health assessment approach by considering effective smart home sensor customization and deployment, data collection, screening, cleaning and filtering, feature computation, normalization and classification, and activity  model training.

\textbf{Research Questions:} \emph{AutoCogniSys} consequently tackles the following key research questions.

$\bullet$ Can we detect simultaneously the periodic rhythms of  both hand gestures and postural activities from wrist-worn ACC sensor signal for diverse population (population with same activity but diverse ways such as walking with walker, stretcher or normally)? If so, how can we incorporate the hand gesture, posture and ambient sensor data streams to help improve the ADLs recognition models?

$\bullet$ How can we exploit and relate the micro-activity features into noise free physiological sensor signals processing to automate cognitive health assessment process? What are the critical roles of clinical survey and technology guided assessment methodologies and their inter-relationships for automating the different intermediate steps of cognitive health assessment process?

To tackle these, we make the following \textbf{key contributions}:

$\bullet$ We employ an extensive signal deconvolution technique that in conjunction with machine learning technique helps facilitate a wrist-worn ACC-based multi-label (hand gestural and postural) activity recognition for diverse population. We then leverage multi-label context sets with ambient and object sensor signals for complex activity recognition based on HDBN model.

$\bullet$ We propose a novel collaborative filter for EDA signal processing by postulating signal as a mixture of three components: \emph{tonic phase, phasic phase} and \emph{motion artifacts}, and employ convex optimization technique for filtering out the motion artifacts. We also propose a novel PPG signal processing technique to filter out the inherent motion artifacts and noises using improved Periodic Moving Average Filtering (PMAF) technique.

$\bullet$  We design and prototype an IoT system consisting of multiple devices (wearable wrist band, IP camera, object and ambient sensors) connected with central hub via WiFi, Ethernet and Bluetooth communication protocols. We collected data from 22 older adults living in a continuing care retirement community center in a very natural setting (IRB \#HP-00064387).

$\bullet$ Finally, we employ statistical and machine learning techniques to jointly correlate the activity performance metrics and stress (EDA and PPG) features that helps achieve max. 93\% of cognitive impairment status detection accuracy. We evaluate \emph{AutoCogniSys} on 5 clinically validated offline assessment tools as ground truth.
\section{Related Works}
\emph{AutoCogniSys} builds on previous works on wearable devices based low-level (postural and hand gestural) activity recognition and their integration with ambient sensors to recognize complex ADLs, the underlying signal processing and applications on cognitive health assessment automation.
\subsection{Wearable Sensor Signal Processing}
Wearable sensors can be two types: physical and physiological. Physical sensors (accelerometer, gyroscope etc.) signal values change over the movements of the sensor devices. Physiological sensors change over physiological condition of body such as EDA changes over stress and PPG changes over heart rate. However, physical movements also impose noises on physiological sensor signals which is called \emph{motion artifacts}.
\subsubsection{Physiological Signal Processing}
A continuous and descrete decomposition of EDA, and time and frequency domain analytics of PPG signal have been investigated before to extract relevant physiological features which were contaminated with noises and motion artifacts \cite{alam16}. \cite{setz10} denoised and classified EDA from cognitive load and stress with accuracy higher than 80\%. Though motion artifacts removal techniques such as exponential smoothing \cite{hern11} and low-pass filters \cite{poh10, hernandez14} provide significant improvement in filtering EDA signals, wavelet transforms offer more sophisticated refinement for any kind of physiological sensors such as electroencephalogram \cite{krish06, zikov02}, electrocardiogram \cite{erc06,alfa08}, and PPG \cite{lee03}. \cite{chen15} proposed a stationary wavelet transform (SWT) based motion artifacts removal technique. `cvxEDA' proposed  a convex optimization technique considering EDA as a mixture of white gaussian noise, tonic and phasic components where white gaussian noise includes motion artifacts and external noises \cite{greco16}. \emph{AutoCogniSys} intelligently combines SWT and `cvxEDA' together to remove noises and motion artifacts from EDA signal. On the other hand, it is more difficult to remove motion artifacts from PPG signal due to its periodicity of nature \cite{wang13}. Researchers proposed different methods such as frequency analytics \cite{garde13,wang13}, statistical analytics \cite{peng14} and digital filter \cite{lee10} to reduce noises and motion artifacts from PPG. \emph{AutoCogniSys} used Periodic Moving Average Filter (PMAF) in this regard \cite{lee07}.
\subsubsection{Physical Sensor Signal Processing}
ACC based hand gesture recognition has been explored by several researchers in past such as discrete hidden markov model \cite{liu10}, artificial neural network \cite{arce11}, weighted naive bayes and dynamic time warping \cite{mace13}. Akl et. al. proposed 18 gesture dictionary based Support Vector Machine (SVM) classifier \cite{akl11}. Wrist-worn ACC based postural activity recognition approach has been proposed using Decision Tree, Random Forest, Support Vector Machines, K-Nearest Neighbors, Naive Bayes and deep neural networks \cite{gj14, wang16}, the accuracy stagnates at 85\% using SVM method \cite{martin16}. However, neither of past works proposed any technique that can provide single body worn ACC sensor-based multiple body contexts recognition nor works efficiently for diverse posture say walking normally, with walker, with double walker or wheel chair. Our proposed 8-hand gesture recognition technique assisted sparse-deconvolution method improves classification performances on both normal and diverse postures. However, we incorporated hand gestures and postures in conjunction with ambient sensors into single-inhabitant HDBN model \cite{alam16b} that provides significant improvement in complex activity recognition.
\subsection{Cognitive Health Assessment}
Smart home environment has been used for providing automated health monitoring and assessment in the ageing population before \cite{dawadi14, gong15, akl15, dawadi15}. `SmartFABER' proposed a non-intrusive sensor network based continuous smart home environmental sensor data acquisition and a novel hybrid statistical and knowledge-based technique to analyz the data to estimate behavioral anomalies for early detection of mild-cognitively impairment \cite{riboni16}. \cite{skubic15} presented an example of unobtrusive, continuous monitoring system for the purpose of assessing early health changes to alert caregivers about the potential signs of health hazards. Though, prior researches proposed a sequence of ambient motion sensor streams as complex activity components in activity based health assessment \cite{dawadi14, gong15, akl15, dawadi15}, we consider inclusion of an wearable wrist-band with in-built ACC sensor to detect hand gesture and posture, augmenting with the ambient sensor readings to help recognize complex activities as well as cognitive health assessment of older adults. Additionally, we propose intelligent use of physiological features of skin through different physiological sensor signals (EDA, PPG) processing in daily activity tasks and incorporate context-awareness for automation of cognitive health assessment that have not been explored before.
\begin{figure}[!htb]
\begin{center}
   \epsfig{file=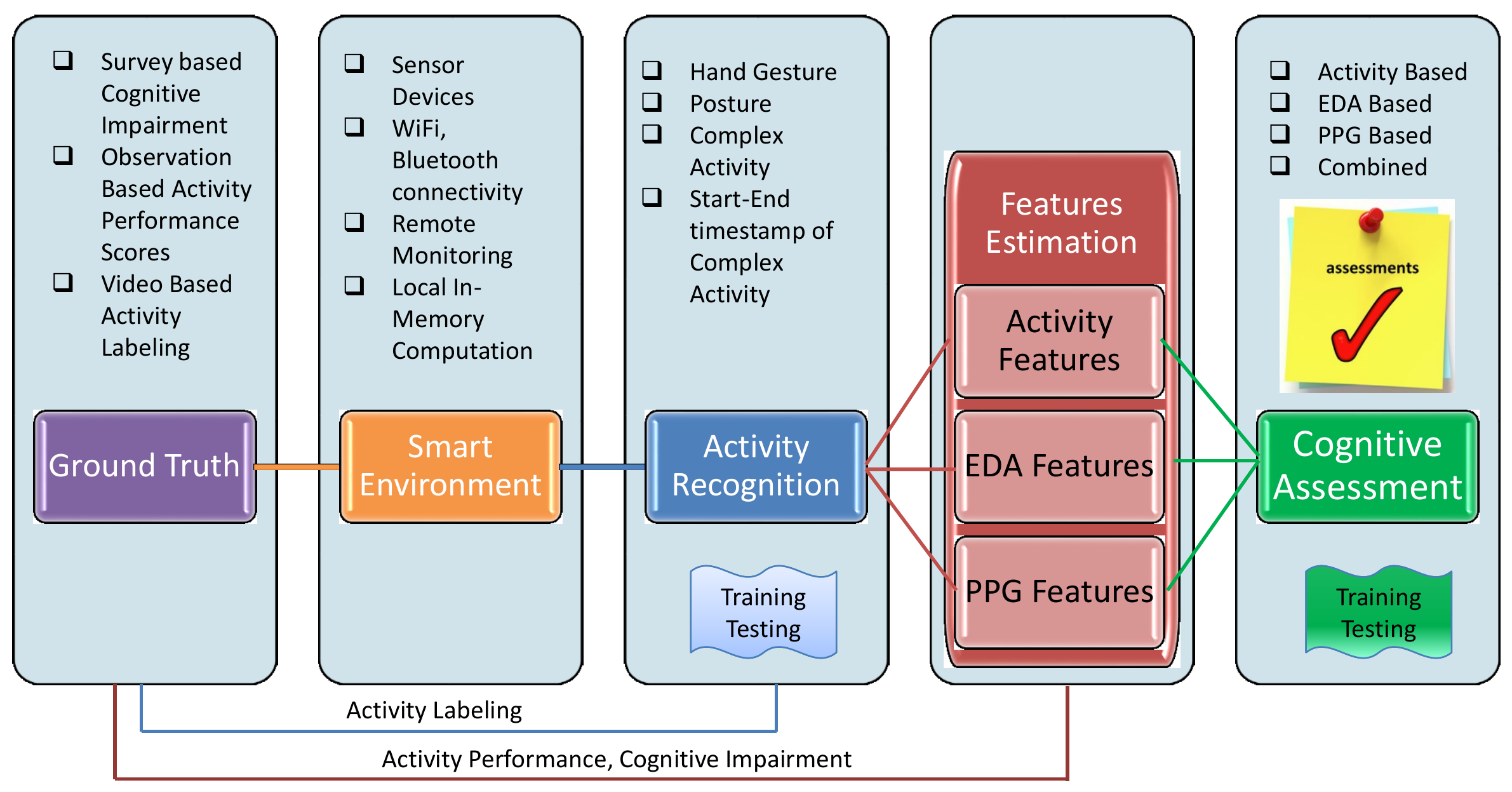,height=1.6in, width=3.5in}
\caption{Overall flow of \emph{AutoCogniSys} pipeline.}
   \label{fig:overview}
\end{center}
\end{figure}
\section{Overall Architecture}
We first investigate existing IoT-based cognitive health care frameworks that covers every aspects of wearable (physical, physiological) and ambient (passive infrared and object) sensor signals computing. \emph{AutoCogniSys} is comprised of three component modules: (i)~sensing, (ii)~processing, and (iii)~analysis. The `sensing' module consists of clinical assessment tools (surveys, observation and clinical backgrounds) and sensing signals (ambient and wearable sensors). `Sensor processing' module is comprised of three sub-modules: a)~clinical assessment feature extraction from assessment tools; b)~ambient sensor feature extraction; and c)~wearable sensor processing (noise removal, segmentation, feature extraction, classification etc.). `Analysis' module is comprised of machine learning and statistical analytics-based score prediction of cognitive impairment. Automation of each module's functionality and inter-intra modular transactions without human interference can be called {\it true automation} of cognitive health assessment. Fig.~\ref{fig:overview} shows the overall flow of \emph{AutoCogniSys} which is discussed in details in the following sections.
\subsection{Demographic Ground Truth Data Collection}
Currently in standard clinical practice and research, the most accurate evaluations of cognitive health assessment are one-to-one observation and supervision tasks/questionnaires for monitoring an individual's functional abilities and behavior \cite{resnick15}. In the first stage of this pilot study, we have investigated current literatures and carefully chosen the clinically proven functional and behavioral health assessment survey tools \cite{resnick15}. On the other hand, to cross check with the survey based evaluations, we have also chosen clinically justified observation based behavioral assessment methods. First, following the resident consent, our clinical research evaluator collects demographic and descriptive data (age, gender, race, ethnicity, marital status, education and medical commodities). She has performed two types of clinical assessments: (1) \emph{Observation based} where the resident's cognition is assessed using the Saint Louis University Mental Status (SLUMS) scale \cite{wai03}. (2) \emph{Survey based} where five widely used and clinically well validated surveys are taken into account: (a) \emph{Yale Physical Activity Survey} \cite{starling99}; (b) \emph{Lawton Instrumental Activities of Daily Living}; (c) \emph{Barthel Index of Activities of Daily Living} \cite{krapp07}; (d) \emph{Geriatric Depression Rating scale} \cite{yesavage82}; and (e) \emph{Zung Self-Rating Anxiety scale} \cite{zung71}.
\subsection{Smart Environment Creation}
For an ideal IoT-based system, instrumenting and deploying it at each participant's natural living environment warrants for assembling a flexible set of hardware and software interfaces to ease the system configuration, setup, and network discovery processes. The sensor system placed in the residences of volunteers needs to meet several specific physiological signals and activity monitoring needs. However, we must confirm that the devices are reliable with potential for re-deployment as well as appear unintimidating to the participants. Inspired by the above requirements, we developed a real testbed IoT system, {\it SenseBox}, by customizing Cloud Engine PogoPlug Mobile base station firmware to integrate with WiFi (connect ambient and object sensors) and Bluetooth (connect wristband) protocol. The smart home components are as follows: (i) PogoPlug base server with a continuous power supply, (ii) 3 binary Passive Infrared sensors in three different rooms (kitchen, livingroom and bedroom) to capture room level occupancy, (iii) 7 binary object sensors attached with closet door, entry door, telephone, broom, laundry basket, trash can and trash box, (iv) three IP cameras in the appropriate positions to collect the ground truth data and (v) an Empatica E4 \cite{empatica} wrist-band (integrated sensors: PPG at 64 Hz, EDA at 4 Hz, Body temperature at 1 Hz and a triaxial ACC at 32 Hz) on the participant's dominating hand.
\section{Activity Recognition}
We aim to detect single wrist-worn ACC sensor based hand gesture and postural activities and insert these into an HDBN graphical model in conjunction with ambient and object sensor values for complex activity recognition. We consider the recognition problem asan activity tupple of $\langle gesture,posture,ambient,object \rangle$. Though, Alam et. al. provides significant performance improvement for single wrist-worn ACC sensor aided 18-hand gesture based postural activity recognition in lab environment \cite{alam17}, it faces some practical challenges in real-time smart environment with older adults due to the diversity of their postures. For example, some older adults use walker, double walking sticks or wheel chair for walking in which cases collecting 18 hand gestures and corresponding postural activities for training requires endless efforts and carefulness. To reduce the complexity of ground truth labeling and later state space explosion for graphical model (HDBN), we propose to use rotational normalization method that can merge some hand-gestures subject to directional differences and forms an 8-hand gesture model. However, our proposed Feature Weight Naive Bayes (FWNB) classifier adds significant improvement on Alam et. al. proposed sparse-deconvolution method as well as recognition in diverse postural environment.
\begin{figure}[!htb]
\begin{center}
   \epsfig{file=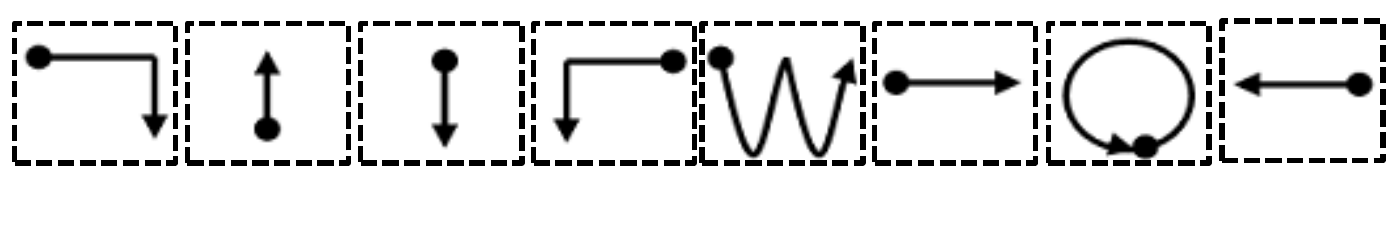,height=0.5in, width=3in}
   \vspace{-.2in}
\caption{8 hand gesture dictionary with direction}
   \label{fig:hand_gestures}
   \vspace{-.2in}
\end{center}
\end{figure}
\subsection{Hand Gesture Recognition}
\label{sec:hand_gesture}
\emph{AutoCogniSys} proposes an 8-gesture dictionary (as shown in Fig. \ref{fig:hand_gestures}) and a Feature Weighted Naive Bayesian (FWNB) framework for building, modeling and recognizing hand gestures. The method comprises of the following steps: (i) \emph{Preprocessing:} wrist-worn ACC sensor provided 3-axis data are passed through 0.4Hz low-pass filter to remove the data drift. (ii) \emph{Rotation normalization:} Normalizing the rotation of hand gestures provides greater accuracy and allows for more realistic, orientation-independent motion. At first, we find the best fit plane of the acceleration vectors thus if the motion lies in a single plane, then the acceleration vectors of a closed shape should on average lie in that main plane. Then, we take all acceleration segments between points of inflection to form one single vector called reference vector that provides us the general direction of user's motion. After that, each vector is normalized relative to the reference vector. This normalization helps remove a lot of hand gestures from prior considered 18 hand gestures resulting a reduced dictionary of 8 gestures. (iii) \emph{Feature Weighted Naive Bayesian model:} Naive Bayes classifier is light-weight and efficient technique for hand gesture recognition. We extract 12 ACC features \cite{alam17} and calculate weight for each feature type based on the similarity of feature measures of the trained gestures (0$<$weight$<$1). While recognizing gestures, the proximity of each feature measure to the average trained feature measure of each gesture type is calculated by a normal distribution. Then, the proximity value is multiplied by the feature weight that was calculated in the training phase.  All of these multiplied values are added together and the system predicts the gesture type with the greatest value as the user gesture. In the learning data points, there should be static postural activities (such as sitting, lying etc.) to avoid unexpected noises over wrist-worn ACC sensors. In the final hand gesture dictionary, we save the reference vector as our signal dictionary.
\subsection{Postural Activity Recognition}
In normal lab environment, wrist-worn ACC sensor signal is a mixture (convolution) of actual hand gesture and postural activity relevant signals \cite{alam17}. \emph{AutoCogniSys} improves the idea by reducing the number of hand gestures and postural activities to 8 (as shown in Fig.\ref{fig:hand_gestures}) using rotation normalization and 4 (walking, sitting, standing and lying). Then, we use sparse-deconvolution method (with 31\% signal reconstruction error) to get Approximately Sparse Factor. The summary of the entire process is stated bellow:

{\it Building Deconvolution Method:} We first consider the wrist-worn ACC sensor signals (3-axis values) as a convolution of hand gesture and postural activity effects and build a deconvolution framework. The deconvolution framework takes a known signal (hand gesture effects) and a equalizer parameter ($\lambda$) as input and provides an Approximately Sparse Factor signal (postural activity effects) as output. For 3-axis ACC signals, we need to learn associated 3 equalizer parameters for each hand gesture. Moreover, each equalizer parameter is involved with 4 postural activities that results a total 96 ($8\times 3\times 4$) equalizer parameters to learn. 

{\it Learning Classification Model:} We use the Approximately Sparse Factor signal to extract 12 statistical features and SVM with sequential machine optimization (SMO) \cite{cao06} for postural activity recognition.

\begin{figure}[!htb]
\begin{center}

   \epsfig{file=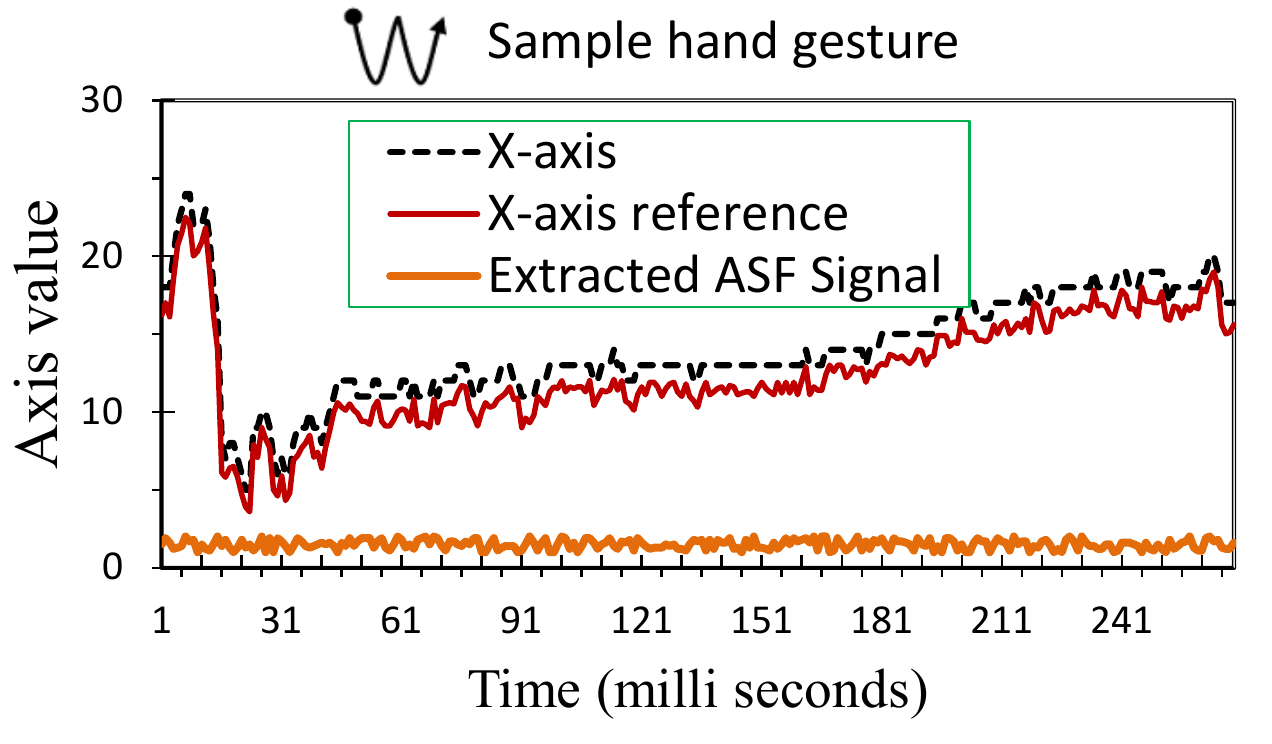,height=1.6in, width=3in}
   \vspace{-.15in}
\caption{Sample deconvolution example of X-axis. The raw x-axis of accelerometer signal, reference vector of the sample gesture and the extracted corresponding ASF signal of walking.}
   \label{fig:deconvolution}
\end{center}
\vspace{-.15in}
\end{figure}

{\it Prediction Model:} After recognizing the hand gestures following the method explained in Sec.~\ref{sec:hand_gesture}, we take the corresponding reference vector as known signal and extract the Approximately Sparse Factor signals incorporating corresponding 3 equalizer parameters ($\lambda$) for the sparse-deconvolution method. Then, we apply feature extraction and prior learned SMO based SVM classifier \cite{cao06} to classify final postural activity. Fig.~\ref{fig:deconvolution} illustrates a single axis example of the deconvolution.

\subsection{Complex Activity Recognition}
We build a HDBN based complex activity recognition framework for single inhabitant scenario smart home environment \cite{alam16b} taking the advantage of detected hand gestural and postural activities along with the ambient and object sensor streams. At first, we obtain instant hand gestural and postural activities from our above proposed models, and additionally motion sensor and object sensor readings from our IoT-system for every time instant generating a 4-hierarchy of HDBN model. Considering the context set $\langle gestural, postural, ambient,object\rangle$ as a hierarchical activity structure (extending two 2-hierarchical HDBN \cite{alam16b}), we build complex activity recognition model for single inhabitant scenario. Finally, we infer the most-likely sequence of complex activities (and their time boundaries), utilizing the well-known Expectation Maximization (EM) algorithm \cite{dempster77} for training and the Viterbi algorithm \cite{forney73} for run-time inference.
\section{Automatic Activity Features Estimation}
The effects of cognitive ability on daily activity performance have been studied before \cite{dawadi14,akl15}. They experimentally and clinically validated that cognitive impairment highly reduces the daily activity performances and this activity performance can be computed as an indicator of cognitive ability status of older adults. The standard activity features refer to completeness of task (TC), sequential task ability (SEQ), interruption avoidance capabilities (INT) etc. In current behavioral science literature, the above activity features carry specific definition based on the sub-tasks involved with a complex activity \cite{dawadi14,akl15}. Completeness of task refers to how many sub-tasks are missed by the participants. Sequential task ability refers to how many sequences of sub-tasks are missed referring the gerontologist defined standard sequences of the sub-task for the particular complex activity. Interruption avoidance capability refers to how many times the participants stop or interleave while doing any sub-task. The final goal of activity features estimation is to provide overall task score. The task score is proportional to the functional ability of participants in performance daily activities. Our behavioral scientist team, comprises with Nursing professor, gerontologist and retirement community caregivers, carefully discus, optimize and choose 87 sub-tasks in total for 13 complex activities.

Each of the sub-task comprises with sequential occurrences of hand gesture and postural activities. However, no researchers ever considered hand gesture for activity features estimation due to complexity of multi-modal wearable and ambient sensors synchronization and multi-label activity classification \cite{dawadi14,akl15}. \emph{AutoCogniSys} exploited single wrist-worn sensor based hand gesture and postural activity recognition, and proposed an activity features (TC, SEQ and INT) estimation method including these two parameters in conjunction with object and ambient sensor features that provide significant improvement of cognitive health assessment of older adults.
\subsection{Machine Learning Based Complex Activity Features Estimation}
In current cognitive health assessment literature, complex activity features can be defined as $\langle TC,SEQ,INT,TS\rangle$. We used supervised method to estimate TC, SEQ and INT, and unsupervised method to estimate TS. We first, formulate the automated scoring as a supervised  machine learning problem in which machine learning algorithms learn a function that maps $\langle${\it hand gesture, posture, object, ambient sensor}$\rangle$ feature set to the direct observation scores. We use bagging ensemble method to learn the mapping function and SMO based SVM \cite{cao06} as base classifier. The learner averages by boostrapping individual numeric predictions to combine the base classifier predictions and generates an output for each data point that corresponds to the highest-probability label. We train three classifiers considering observation as ground truth for TC, SEQ and INT scores and test on the testing dataset. We derive unsupervised scores using dimensionality reduction technique for each feature set. First, we take all features of each activity, apply optimal discriminant analysis technique as a dimensionality reduction process \cite{zhang09} and reduce the feature sets into single dimensional value which represents the automated task completeness scores of the particular user activity. A min-max normalization is applied that provides us a uniform range of the variables using $
z_i=\frac{x_i-min(x)}{max(x)-min(x)}$ equation where $x=\{x1,\ldots,x_n\}$ and $z_i$ is $i^{th}$ normalized data. The final single dimensional score represents machine learning based TS score.
\section{Physiological Sensor Signals Processing}
The autonomic nervous system (ANS) restrains the body's physiological activities including the heart rate, skin gland secretion, blood pressure, and respiration. The ANS is divided into sympathetic (SNS) and parasympathetic (PNS) branches. While SNS actuates the body's resources for action under arousal conditions, PNS attenuates the body to help regain the steady state. Mental arousal (say stress, anxiety etc.) activates the sweat gland causing the increment and reduction of  Skin Conductance on SNS and PNS physiological conditions respectively. However, Instant Heart Rate also has similar effect on SNS and PNS physiological condtions i.e., a higher value of heart rate is the effect of SNS and lower value is the outcome of PNS. EDA and PPG sensors are widely used to estimate the instant value of skin conductance and heart rate respectively \cite{alam16}.
\subsection{EDA Sensor Signal Processing}
EDA is the property of the human body that causes continuous variation in the electrical characteristics of the skin which varies with the state of sweat glands in the skin. There are three types of arousal: \emph{cognitive, affective and physical}. \emph{Cognitive} arousal occurs when a person tries to solve any problem using her cognitive ability. \emph{Affective} arousal occurs when a person is worried, frightened or angry either doing daily activities or in resting position. On the other hand, \emph{physical} arousal is related to the brain command to move bodily parts which is imposed on the total arousal as an artifact, called \emph{motion artifact}. However, there are always some noises due to the weather conditions (temperature, humidity etc.) and device motion. This \emph{motion artifact} can be the prime cause of signal contamination of physiological outcomes while performing daily activities which must be removed. \emph{AutoCogniSys} proposes an EDA sensor signal processing method consists of three steps: (i) noise and motion artifacts removal, (ii) separation of tonic component and phasic component (explained later) from contamination free EDA signal and (iii) feature extraction on the response window.
\subsubsection{Motion Artifacts Removal}
\begin{figure}[!htb]
\begin{center}
\vspace{-.1in}
   \epsfig{file=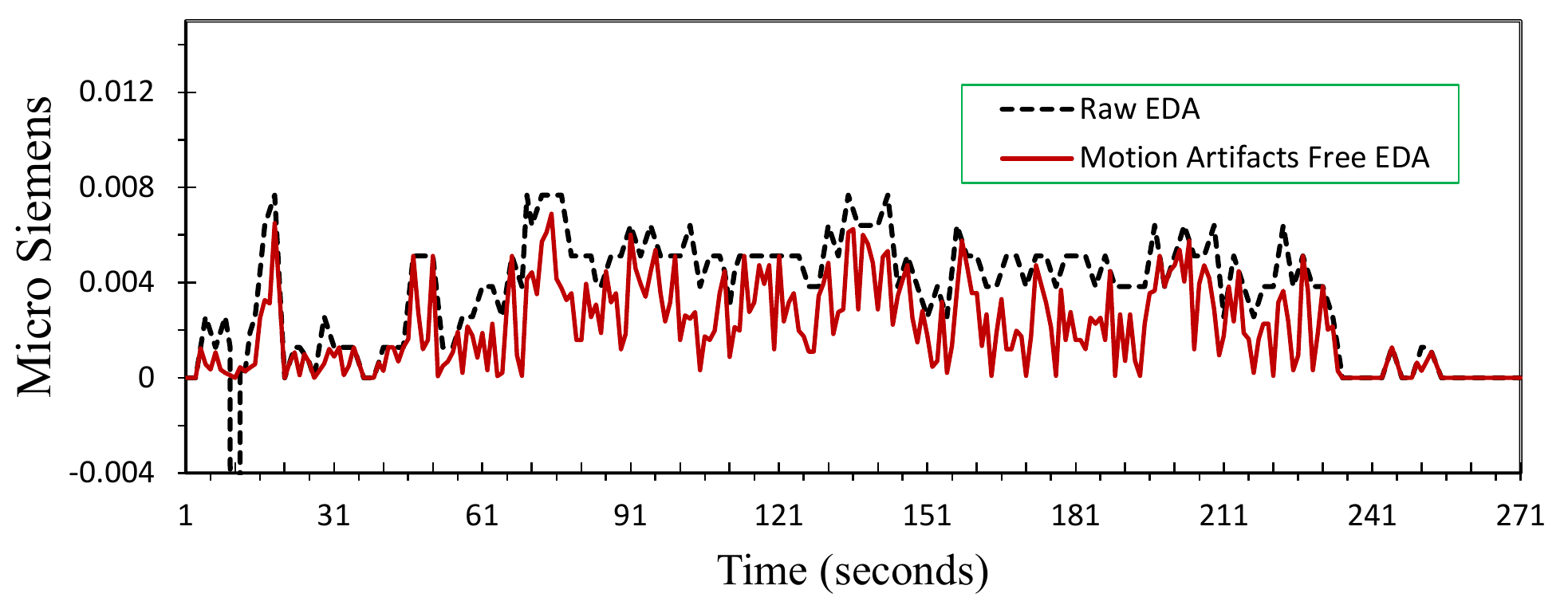,height=1.6in, width=3.5in}
\caption{Dashed line represents noisy EDA signal and solid red line represents \emph{AutoCogniSys} proposed motion artifact free EDA signal}
   \label{fig:eda_artifact_removal}
\end{center}
\end{figure}

There are many types of motion artifacts but the unsual steep rise is the mostly occured ones associated with EDA signal while performing daily activities \cite{edel67}. We use well-known steep rising noises reduction technique, SWT \cite{chen15}. We first consider EDA signal as a mixture of a slow variant tonic and fast variant phasic component, i.e., SWT coefficient is modeled as a mixture of two Gaussian components, phasic (close to zero valued signal) and tonic (high rising signal). After expanding EDA signal into multiple levels of scaling and wavelet coefficients, we choose adaptively a threshold limit at each level based on the statistical estimation of the wavelet coefficients' distribution, and employ that on the wavelet coefficients of all levels. Finally, an inverse wavelet transform technique is applied to the thresholded wavelet coefficients to obtain the artifacts free EDA signal. Fig~.\ref{fig:eda_artifact_removal} shows a sample of raw and motion artifacts free EDA signal.

\subsubsection{Convex Optimization Technique to EDA Deconvolution}
After the motion artifact removal, we consider EDA as the sum of three components for $N$ sample: a slow tonic driver ($t$), fast (compact, bursty) non-negative sparse phasic driver ($r$) and a reminder error term ($\epsilon_r$).
\begin{equation}
\label{eq:eda_signal}
y = t + r + \epsilon_r
\end{equation}
This additive error $\epsilon_r$ is a White Gaussian Noise. The central problem associated with the deconvolution method is to get tonic $t$ component from the above equation. \cite{greco16} showed that EDA signal deconvolution (separation of tonic, phasic and noise terms from EDA signal) is a quadratic optimization problem and defined tonic component as follows:
\begin{equation}
\label{eq:tonic}
t = Bl + Cd,
\end{equation}
where $B$ is a tall matrix whose columns are cubic $B$-spline basis functions, $l$ is the vector of spline coefficients, $C$ is a $N\times 2$ matrix, $d$ is a $2\times 1$ vector with the offset and slope coefficients for the linear trend. The above equation is subject to the following optimization problem,
\begin{eqnarray}
minimize \frac{1}{2} {||Mq + Bl + Cd- y||}^2_2 +\alpha {||Aq||}_1 + \frac{\lambda}{2} {||l||}^2_2\\
subject\;to\; Aq \geq 0\nonumber
\end{eqnarray}
where $M$ and $A$ are tridiagonal matrices and $q$ is an auxiliary variable. After solving the above equation, we can get the optimal values for $\{q,l,d\}$ that can be used to obtain tonic component from the equation~\ref{eq:tonic}. The reminder of the equation~\ref{eq:eda_signal} ($r+\epsilon_r$) is considered as a mixture of White Gaussian Noise ($\epsilon_r$) and a fast variant phasic component ($r$). We employ butterworth low-pass filter (5Hz) and hanning smoothing with window size 4 (optimal) to remove $\epsilon_r$ from phasic component ($r$).
\subsection{PPG Signal Processing}
PPG is used mainly for measuring the oxygen saturation in the blood and blood volume changes in skin. An ideal PPG signal processing must contain the following steps: noise and motion artifacts removal, heart rate detection, heart rate variability estimation and feature extraction.
\subsubsection{PPG Signal Noise and Motion Artifacts Removal}
Similar to EDA signal, PPG signal is also contaminated with motion artifacts and noises. However, unlike EDA signal, PPG produce quasiperiodicity in a time series spectrum \cite{mete30}. We use Periodic Moving Average Filter (PMAF) to remove motion artifacts and noises \cite{lee07}. We first segment the PPG signal on periodic boundaries and then average the $m^{th}$ samples of each period. After filtering the input PPG signal with a 5-Hz $8^{th}$-order Butterworth low-pass filter, we estimate the maximum and minimum value of each period. The mean of each period are obtained from the maximum and minimum values applying the zero crossing method. These points of the means help determine the boundaries of each period. Then, interpolation or decimation is performed to ensure that each period had the same number of samples \cite{lee07}. 
\subsubsection{Heart Rate and Heart Rate Variability Estimation}
\begin{figure}[!htb]
\vspace{-.1in}
\begin{center}
   \epsfig{file=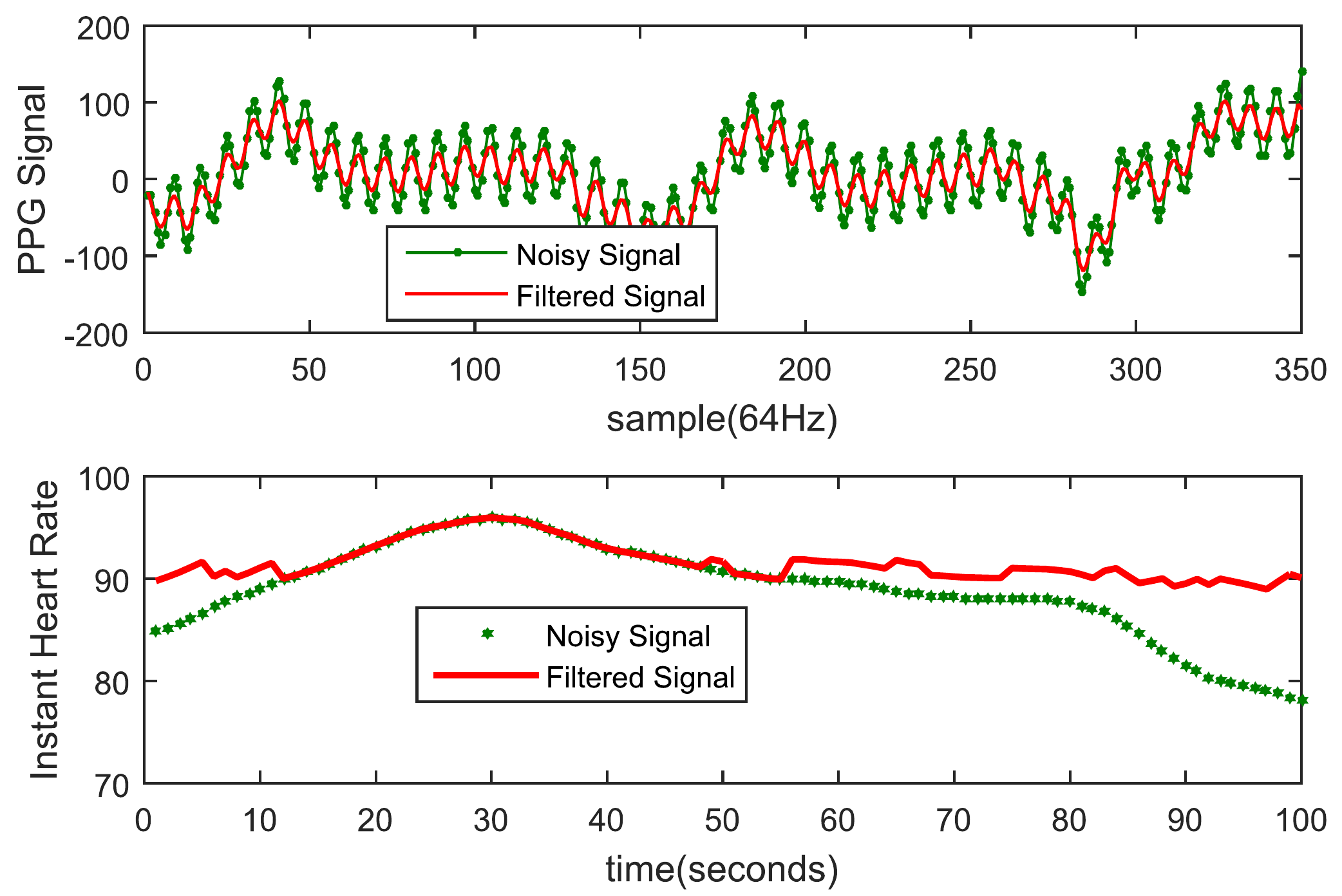,height=1.4in, width=3.5in}
   \vspace{-.15in}
\caption{Top figure illustrates the noisy signal (dotted line) and filtered signal from PPG sensor based on our filtering method. Bottom figure illustrates instant heart rate calculated from noisy signal (dotted line) and filtered signal}
   \label{fig:ppg_artifact_removal}
\end{center}
\vspace{-.15in}
\end{figure}

We first apply PMAF on PPG signal to remove noises and motion artifacts, refine PPG by smoothing the signal using 1-dimensional Gaussian Filter and Convolution, calculate first derivative of the convoluted signal and finally find the differences between two consecutive peak values which is called HRV \cite{sel08}. The occurrences of total peak values (R-peak or beat) in each minute is called Heart Rate (HR) with an unit of Beat Per Minute. The signal value property of HRV and HR are inversely proportional which means the mental arousal that increases HR should decrease HRV in the time segment window. Fig~\ref{fig:ppg_artifact_removal} shows a sample of the noisy and filtered PPG signal and their corresponding Instant Heart Rate.
\subsection{Physiological Sensor Signal Feature Extraction}
Using the above mentioned methods, we removed the noises and motion artifacts from EDA and PPG signals and generated two time series signal from EDA (tonic and phasic components) and one time series signal from PPG (HRV). Then, we segment each of the time series signal based on our prior detected complex activities such that each response window starts and ends with the starting and ending points of each complex activity. We extract 7 statistical time-series features for EDA (as shown in Table~\ref{tab:eda_features}) and 8 features for HRV (Table~\ref{tab:hrv_features}) within the response window).

\begin{table}[!t]
\begin{center}

\renewcommand{\arraystretch}{1}
\caption{EDA Features Within The Response Window}
\begin{scriptsize}

\label{tab:eda_features}
\begin{tabular}{|c|l|}
\hline
\bfseries Features& \bfseries Description\\
\hline
nSCR & Number of SCRs within response window (wrw)\\
\hline
Latency & Response latency of first significant SCR wrw\\
\hline
AmpSum & Sum of SCR-amplitudes of significant SCRs wrw\\
\hline
SCR & Average phasic driver wrw\\
\hline
ISCR & Area (i.e. time integral) of phasic driver wrw\\
\hline
PhasicMax & Maximum value of phasic activity wrw\\
\hline
Tonic & Mean tonic activity wrw\\
\hline
\end{tabular}
\end{scriptsize}
\end{center}
\end{table}

\begin{table}[!t]
  \begin{center}
\renewcommand{\arraystretch}{1}
\vspace{-.3in}
\caption{Heart Rate Variability features}
\label{tab:hrv_features}
\begin{scriptsize}
\begin{tabular}{|c|l|}

\hline
\bfseries Feature& \bfseries Description\\
\hline
$\overline{RR}$&Mean RR intervals\\
\hline
SDNN&Standard deviation of RR intervals\\
\hline
SDSD&Std of successive RR interval differences\\
\hline
RMSSD&Root mean square of successive differences\\
\hline
NN50&\#successive intervals differing more than 50 ms\\
\hline
pNN50&relative amount of NN50\\
\hline
HRVTI&Total number of RR intervals/height of the histogram\\
\hline
TINN&Width of RR histogram through triangular interpolation\\
\hline
\end{tabular}
\end{scriptsize}
  \end{center}
\end{table}
\section{Experimental Evaluation}
In this section, we explain our data collection, available benchmark dataset, baseline methods and evaluation.
\subsection{Datasets and Baseline Methods}
We validate and compare \emph{AutoCogniSys} with baseline methods on both publicly available and our collected datasets.
\subsubsection{RCC Dataset: Collection and Ground Truth Annotation}
For collecting Retirement Community Center Dataset (RCC Dataset), we recruited 22 participants (19 females and 3 males) with age range from 77-93 (mean 85.5, std 3.92) in a continuing care retirement community with the appropriate institutional IRB approval and signed consent. The gender diversity in the recruited participants reflects the gender distribution (85\% female and 15\% male) in the retirement community facility. A trained gerontology graduate student evaluator completes surveys with participants to fill out the surveys. Participants are given a wrist band to wear on their dominant hand, and concurrently another trained IT graduate student have the IoT system setup in participants' own living environment (setup time 15-30 minutes). The participants are instructed to perform 13 \emph{complex ADLs}. Another project member remotely monitors the sensor readings, videos and system failure status. The entire session lasts from 2-4 hours of time depending on participants' physical and cognitive ability.

We follow the standard protocol to annotate demographics and activities mentioned in the IRB. Two graduate students are engaged to annotate activities (postural, gestural and complex activity) whereas the observed activity performances are computed by the evaluator. Two more graduate students are engaged to validate the annotations on the videos. In overall, we are able to annotate 13 complex activities (total 291 samples) labeling for each participant; 8 hand gestures (total 43561 samples) and 4 postural activities (total 43561 samples) labeling. Annotation of postural and complex activities outcomes no difficulties from recorded videos. However, annotation of hand-gestures is extremely difficult in our scenario. We used video based hand tracker that can track and sketch wrist movements from a video episode \cite{hugo14}. This sketching can help us significantly to identify which particular hand gesture is being performed in the time segment.
\subsubsection{EES Datasets: EDA and PPG Sensor Datasets}
We used Eight-Emotion Sentics (EES) dataset to validate \emph{AutoCogniSys} proposed physiological signal processing approaches \cite{picard01}. The dataset consists of measurements of four physiological signals (PPG/Blood Volume Pulse, electromyogram, respiration and Skin Conductance/EDA) and eight affective states (neutral, anger, hate, grief, love, romantic love, joy, and reverence). The study was taken once a day in a session lasting around 25 minutes for 20 days of recordings from an individual participant. We consider only PPG and EDA for all of the affective states in our study.
\subsubsection{Baseline Methods}
Though no frameworks ever combined all modalities together into real-time automated cognitive health assessment, we evaluate \emph{AutoCogniSys} performance by comparing the performances of its components individually with upto date relevant works. For hand gesture and postural activity recognition, we consider \cite{alam17} proposed method as baseline. For complex activity recognition, we compare our hand gesture and postural activity classifiers aided HDBN model with three-level Dynamic Bayesian Network \cite{zhu12} framework. For activity performance estimation, activity performance based cognitive health assessment; and EDA and PPG based cognitive health assessment, we have considered \cite{alam16} proposed method as baseline.
\subsection{Activity Recognition Evaluation}
The standard definition for \emph{accuracy} in any classification problem is $\frac{TP+TN}{TP+TN+FP+FN}$ where $TP,TN,FP$ and $FN$ are defined as true positive, true negative, false positive and false negative. For complex activity recognition evaluation, we additionally consider \emph{start/end duration error} as performance metric that can be explained as follows: consider that the true duration of ``cooking'' is 30 minutes (10:05 AM - 10:35 AM) and our algorithm predicts 29 minutes (10.10 - to 10.39 AM). Then, the start/end duration error is 9 minutes ($|$5 minutes delayed start$|$ + $|$4 minutes hastened end$|$), in an overall error of e.g., 30\% (9/30=0.3). We measure cross-participant accuracy using leave-two-participants-out method for performance metrics, i.e., we take out two of the participants' data points from the entire dataset, train our proposed classification models, test the model accuracy on the two left-out participants relevant data points, and continue the process for entire dataset.

\begin{figure*}[!htb]
\begin{minipage}{0.45\textwidth}
\begin{center}
   \epsfig{file=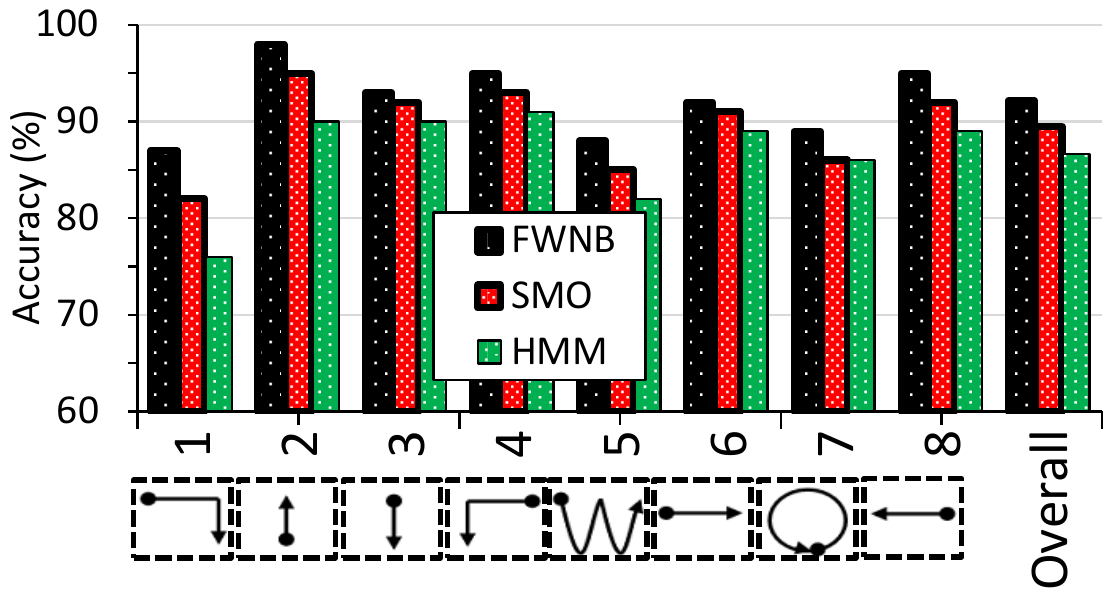,height=1.6in, width=3in}
\caption{Feature Weighted Naive Bayes (FWNB) classification accuracy comparisons with baseline approaches (graphical signatures of all hand gestures are shown).}
   \label{fig:hand_gesture_accuracy}
\end{center}
\end{minipage}
\begin{minipage}{0.29\textwidth}
\begin{center}
\vspace{-.12in}
   \epsfig{file=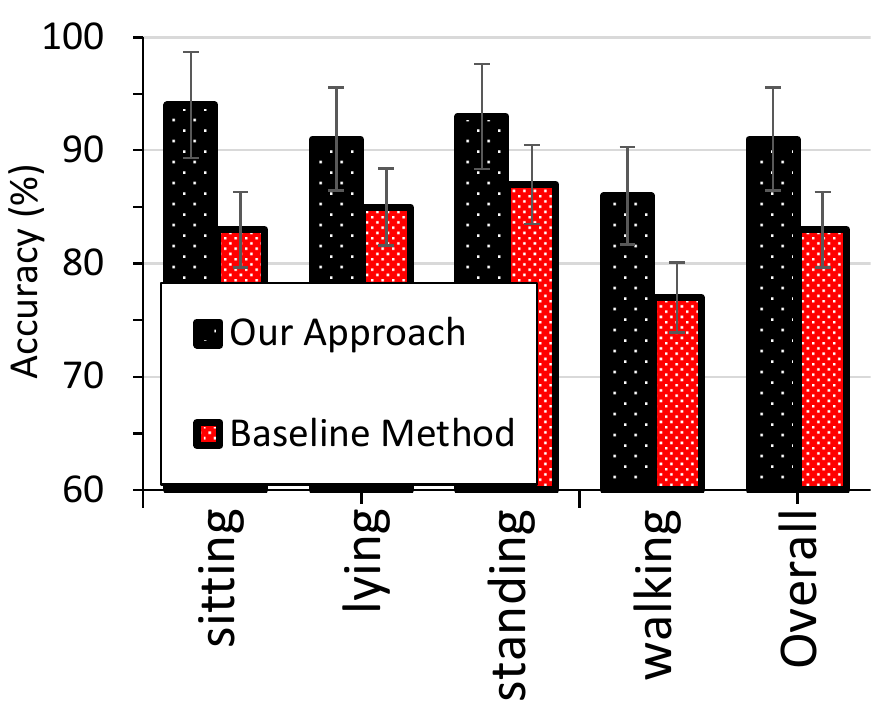,height=1.6in, width=2.1in}
\caption{4-class postural level activity recognition performance and comparisons with baseline method}
   \label{fig:posture_accuracy_normal}
\end{center}
\end{minipage}
\begin{minipage}{0.25\textwidth}
 \begin{center}
 \vspace{-.12in}
   \epsfig{file=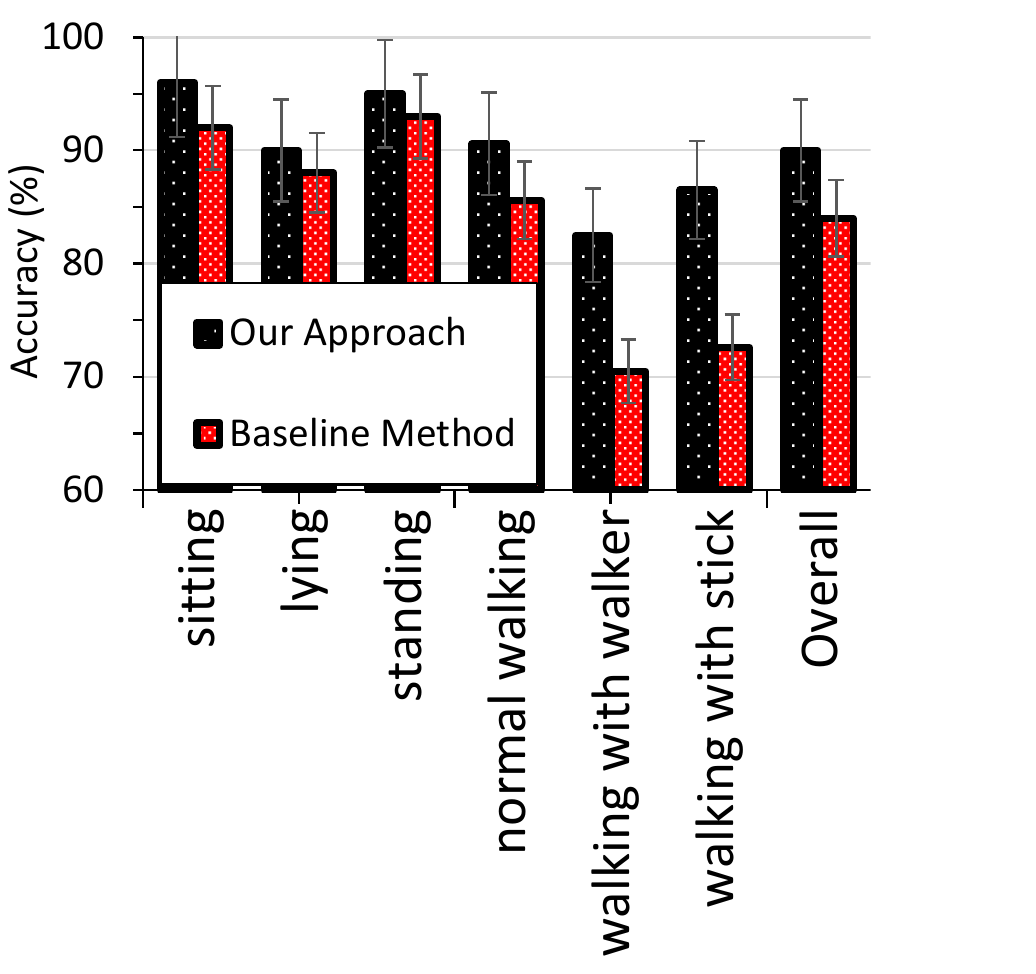,height=1.6in, width=2.1in}
\caption{6-class diverse postural activity recognition framework accuracy comparisons with the baseline approach.}
   \label{fig:posture_accuracy_extended}
\end{center}
 \end{minipage}
\end{figure*}

Fig~\ref{fig:hand_gesture_accuracy} displays Feature Weighted Naive Bayes (FWNB) based the 8-hand gestural activity recognition accuracies comparisons with the baseline methods which clearly depicts the outperformance of our method (5\% improvement) with an overall accuracy of 92\% (FP rate 6.7\%) in RCC dataset. For postural activity recognition, dataset achieving 91\% postural activity recognition accuracy (FP rate 9.5\%) which outperforms the baseline approach significantly (8\% improvement). Now, we expand the postural activities for RCC datasets into 3 diverse `walking' postures: `normal walking', `walking with walker', `walking with single stick' and the accuracy goes down to 88\% (FP 7.9\%). Fig.~\ref{fig:posture_accuracy_normal} and Fig.~\ref{fig:posture_accuracy_extended} illustrate 4-class postural and extended 6-class postural classifier accuracies respectively which clearly posit that \emph{AutoCogniSys} outperforms in each case of postural activities as well as overall performances (8\% and 7\% improvement respectively).

\begin{figure}  [!htb]
  \begin{minipage}{0.15\textwidth}
 \begin{center}
   \epsfig{file=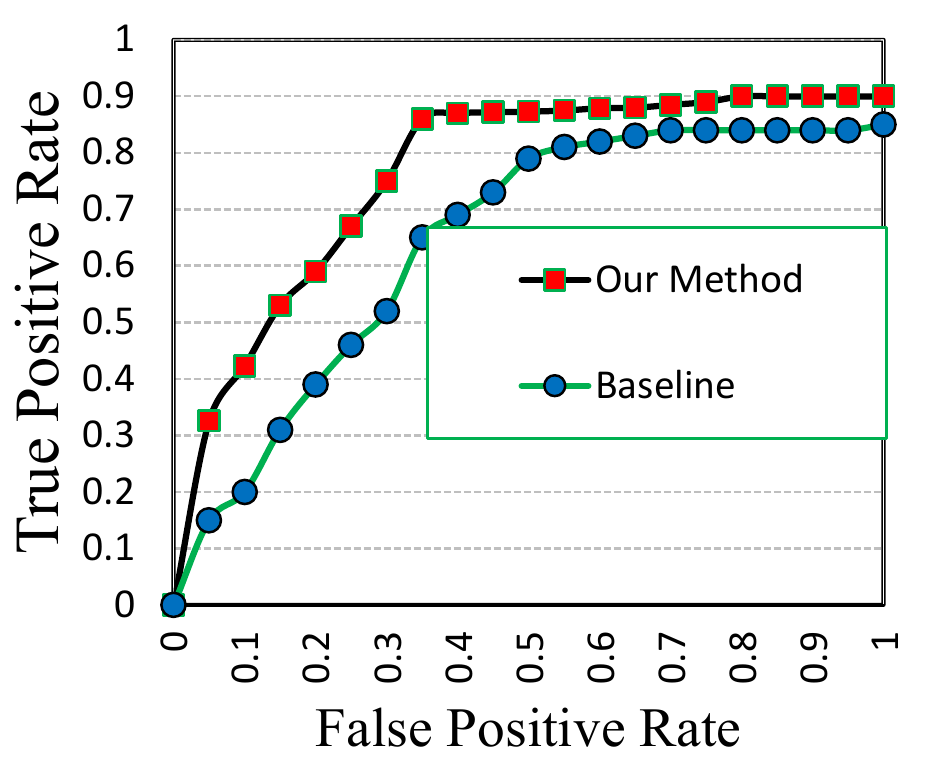,height=1.4in, width=1.1in}
\caption{ROC curve for complex activity recognition}
   \label{fig:complex_activity_roc}
\end{center}
\end{minipage}
\begin{minipage}{0.33\textwidth}
\begin{center}

   \epsfig{file=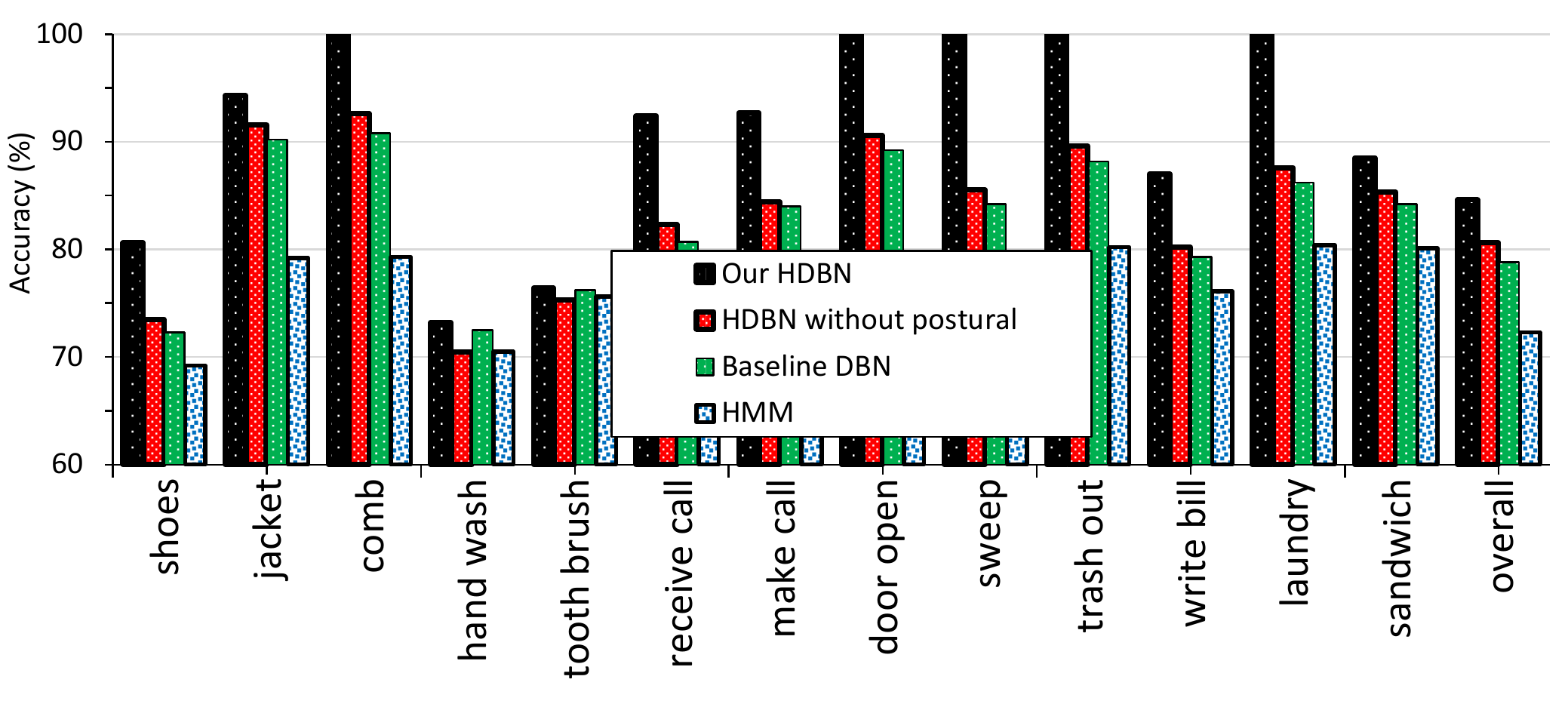,height=1.4in, width=2.3in}
\caption{Complex ADLs recognition accuracy improvement and comparison with baseline \cite{zhu12} and HMM based method}
   \label{fig:complex_activity_accuracy}
\end{center}

\end{minipage}
\end{figure}

For complex activity classification, we choose RCC dataset to train our HDBN model. Our leave-two-participants out method results an accuracy of 85\% (FP Rate 3.6\%, precision 84.2\%, recall 84.5\%, ROC Area 98.2\%) with a start/end duration error of 9.7\%. We run the entire evaluation for baseline complex activity recognition algorithm too achieving an overall accuracy of 78\% (FP Rate 5.2\%, precision 79.6\%, recall 78.5\%, ROC Area 82.7\%) which is clearly lower performed method than our approach. Fig. \ref{fig:complex_activity_roc} and Fig~\ref{fig:complex_activity_accuracy} illustrate the ROC curve and each complex activity recognition accuracy comparisons with baseline method which depict the outperformance of our framework over baseline methods (7\% improvement). Fig~\ref{fig:complex_activity_accuracy} also shows that inclusion of postural activity improves the final complex activity recognition (4\% improvement).

\subsection{Quantification of Performance Score}
To characterize both the qualitative and quantitative health assessment performance scores, we start with four different feature groups ranging from both functional and physiological health measures: (i) observation based activity features, (ii) automatic activity performance features, (iii) EDA features and (iv) PPG features.

In \emph{observation based activity features}, we design a complex activity set comprised of multiple subtasks which are involved with task {\it interruption, completion and sequencing}. Participants are instructed to perform the complex activities while the trained evaluator observed the aforementioned functional activity performance measures. Each incorrect attempt of performance measure will be assigned one point thus higher score reflects lower performance of functional activities \cite{dawadi14}. We first detect hand gesture and postural activities. Then, we feed the low-level activity contexts (gestural and postural) combined with ambient contexts (object and ambient motion sensor readings) into HDBN for single inhabitant model \cite{alam16b} to recognize complex activities. The complex activity recognition framework provides both activity labels and activity window (start-end points). Then, we extract features of object sensor, ambient sensor, gestural activity and postural activity events for each activity window. The features are number of occurrences, mean number of occurrences, consecutive 1, 2, 3, $\ldots$ 20 occurrences, top 10, 20, 30, $\ldots$, 90 percentile etc (29 features in total). In \emph{physiological features} we first detect 13 complex activities using HDBN algorithm which provides activity labels and activity window (start-end points), apply noise reduction, motion artifacts removal, extract 7 EDA features and 8 HRV features for each activity and take the mean of them over time (minutes) to get 15 (7+8) complex activity physiological features set for each participant. In summary, we extract 3 observation based activity features, 29 automatic activity performance features, 7 EDA features and 8 HRV features.
\subsection{Physiological Signal Processing Performance Evaluation}
Standard evaluation technique should use both experimental and publicly available datasets to confirm the outperformance of the novel approaches. We first evaluate our physiological signal processing techniques using a publicly available dataset (EES Dataset \cite{picard01}) to detect 8 human emotions. Then, in next section, we evaluate our methods in assessing cognitive health status of older adults using RCC dataset.

\begin{figure}[!htb]
\begin{minipage}{0.24\textwidth}
\begin{center}
   \epsfig{file=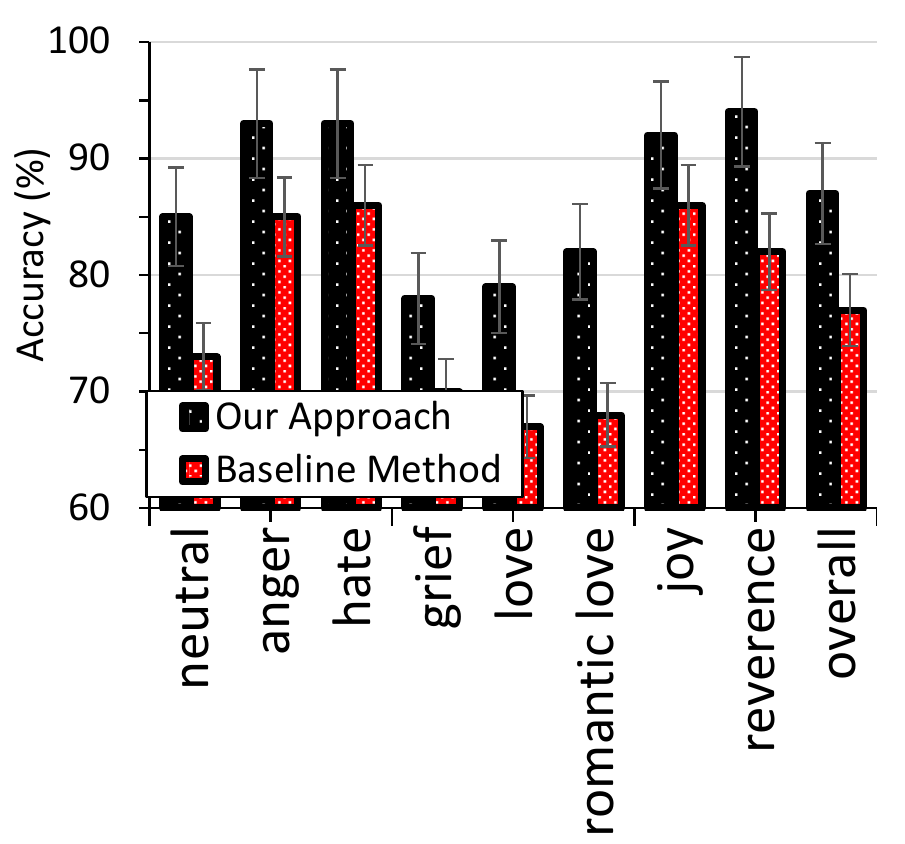,height=1.2in, width=1.8in}
\caption{(EES Databaset) EDA features based Eight Emotion classification accuracy comparisons with baseline method}
   \label{fig:ees_eda}
\end{center}
\end{minipage}
\begin{minipage}{0.23\textwidth}
\begin{center}
   \epsfig{file=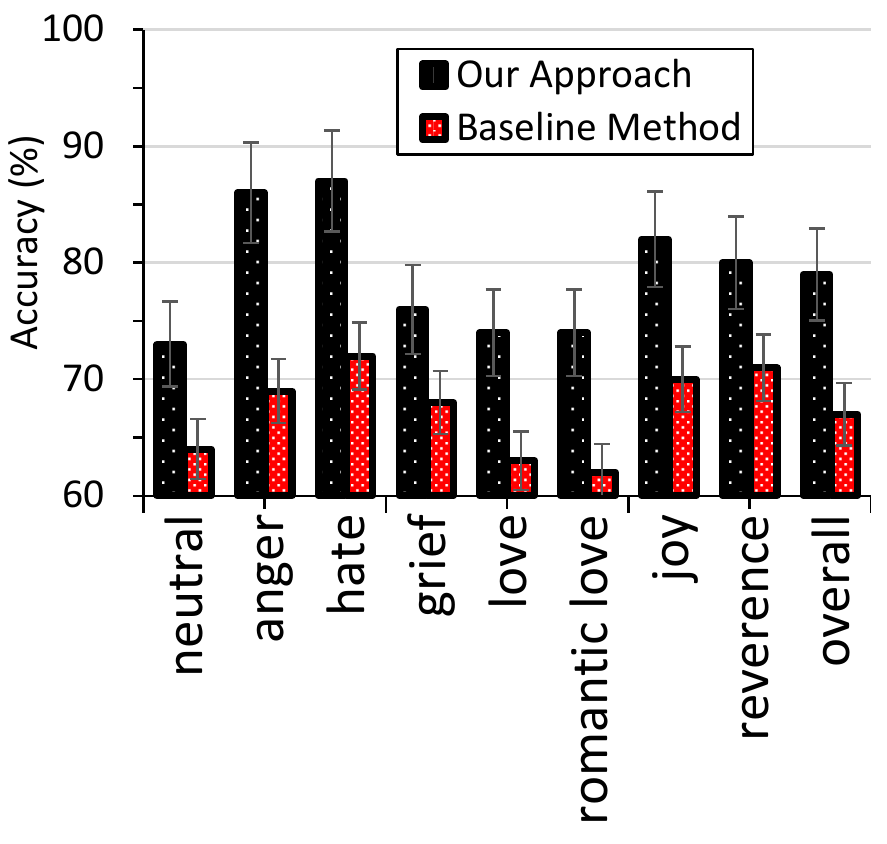,height=1.2in, width=1.7in}
\caption{(EES Dataset) PPG features based 8-Emotion classification accuracy comparisons with baseline method}
   \label{fig:ees_ppg}
\end{center}
\end{minipage}

\end{figure}

For EDA, we first apply SWT method to remove motion artifacts and noises. Then, we use cvxEDA method to separate tonic and phasic components of EDA. Then, we extract 7 EDA features on a sliding window of 4 seconds. Finally, we feed the 7 EDA features into a SMO based SVM algorithm \cite{cao06}. We use 10-fold cross validation to classify eight emotions achieving 87\% of overall accuracy (FP rate 6\%). For PPG, we first apply our proposed PMAF based noises and motion artifacts removal technique. Then, we calculate HRV and perform time-domain feature extraction to extract 8 HRV features on a sliding window of 4 seconds. We feed these features into a SMO based SVM algorithm \cite{cao06}. Our 10-fold cross validation shows accuracy of 79\% (FP rate 11.5\%) of detecting 8 emotions on EES Dataset. Fig. \ref{fig:ees_eda} and Fig. \ref{fig:ees_ppg} clearly depict that \emph{AutoCogniSys} proposed EDA and PPG signal processing techniques significantly improve the accuracy over the baseline \cite{alam16} method (10\% and 12\% improvement).

\subsection{Evaluation of Performance Scores}
The feature subsets used in the experimentation for observation and survey based clinical assessments and technology guided physiological and activity initiated health assessments are depicted in Table~\ref{tab:feature_subset}. From our 6 demographics surveys, we find significant distributions in terms of cognition only for SLUMS Score (S-Score). Based on that, we divide our participants pool into three groups: \emph{Not Cognitively Impaired (NCI), Mild Cognitively Impaired (MCI) and Cognitively Impaired (CI)} where the number of participants are $5$, $7$ and $10$ respectively.
\begin{table}[!t]
\begin{scriptsize}

{\centering 
\renewcommand{\arraystretch}{.6}
\caption{Feature Subsets}
\label{tab:feature_subset}
\begin{tabular}{|l|L{5.5cm}|}
\hline
\bfseries Feature& \bfseries Description\\
\hline
Observation & Task Completeness (TC), Sequencing (SEQ), Interruptions (INT)\\
\hline
Survey & SLUMS Score (S-Score), ZUNG Score (Z-Score), IADL Score (I-Score), Yale Score (YPAS), Barthel Score (B-Score), GDS Score (G-Score)\\
\hline
EDA  and HRV & 7 and 8 Features\\
\hline
Activity Performance& Supervised (TC, SEQ, INT), Unsupervised\\
\hline
Arousal& EDA and HRV features of each complex activity window\\
\hline

\end{tabular}
}
\end{scriptsize}
\end{table}

\begin{figure}[!htb]
\begin{center}
   \epsfig{file=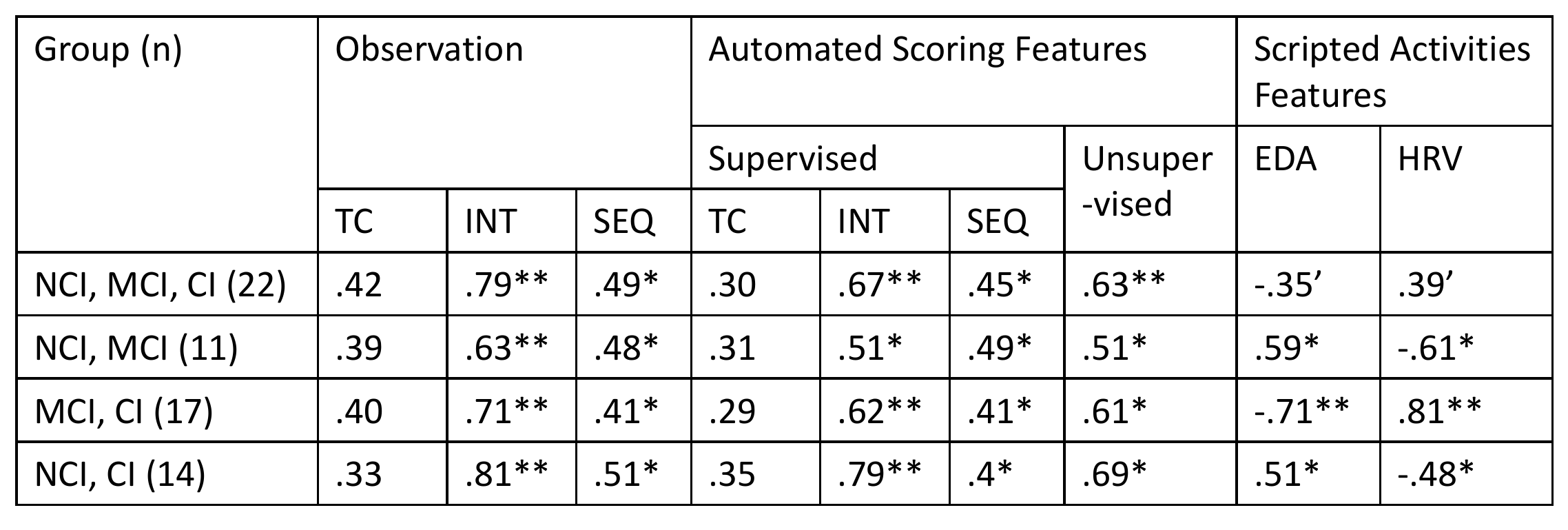,height=1in, width=3.3in}
\caption{\emph{AutoCogniSys} Proposed Method Based Group Correlation analysis ( $r-value$) NCI, MCI and CI represent not cognitive, mild cognitive and cognitively impaired group of population. TC, INT, SEQ, EDA and HRV represent task completeness, interruption scores, sequencing scores, electrodermal activity features and heart rate variability features}
   \label{fig:group_correlation}
\end{center}
\vspace{-.2in}
\end{figure}
\begin{figure}[!htb]
\begin{center}
   \epsfig{file=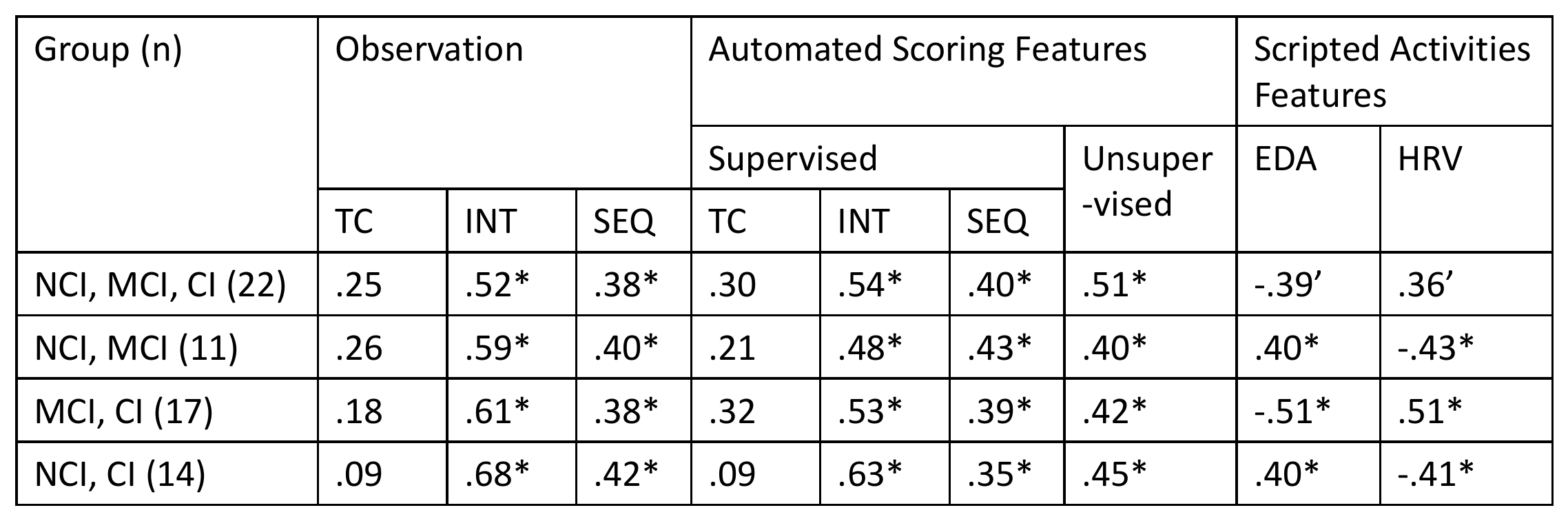,height=1in, width=3.3in}
\caption{Baseline \cite{alam16} method based Group Correlation analysis ( $r-value$)}
   \label{fig:group_correlation_baseline}
   \vspace{-.25in}
\end{center}
\end{figure}

\subsection{Statistical Correlation Analysis of Cognitive Health}
We used Pearson correlation coefficients with significance on $p<0.05$* for individual feature and partial correlation coefficients with significance on $p<0.005$** for group of features correlation analysis. Fig. \ref{fig:group_correlation} and Fig. \ref{fig:group_correlation_baseline} show the group correlation analysis results based on \emph{AutoCogniSys} proposed framework and baseline \cite{alam16} framework respectively. It can be clearly depicted that our proposed framework improves the correlation with the ground truths.
\subsection{Machine Learning Classification of Cognitive Health}
We evaluate using machine learning classifiers to predict cognitive status of older adults using both individual modalities and combined features. We use leave-two-participants out method to train and test classification accuracy.

We first choose the individual activity features (machine learning method based interruption scores, sequencing scores, unsupervised scores) and their combined features to train and test cognitive impairment status classification for SMO based SVM algorithm \cite{cao06}. The classification accuracies are 72\%, 69\%, 76\% and 83\% respectively. Then we consider 7 EDA-activity features and 8 HRV-activity features individually in training and testing phase of SMO based SVM algorithm \cite{cao06} resulting 85\% and 80\% accuracy respectively.

\begin{figure}[!htb]
\begin{minipage}{0.24\textwidth}
\begin{center}
   \epsfig{file=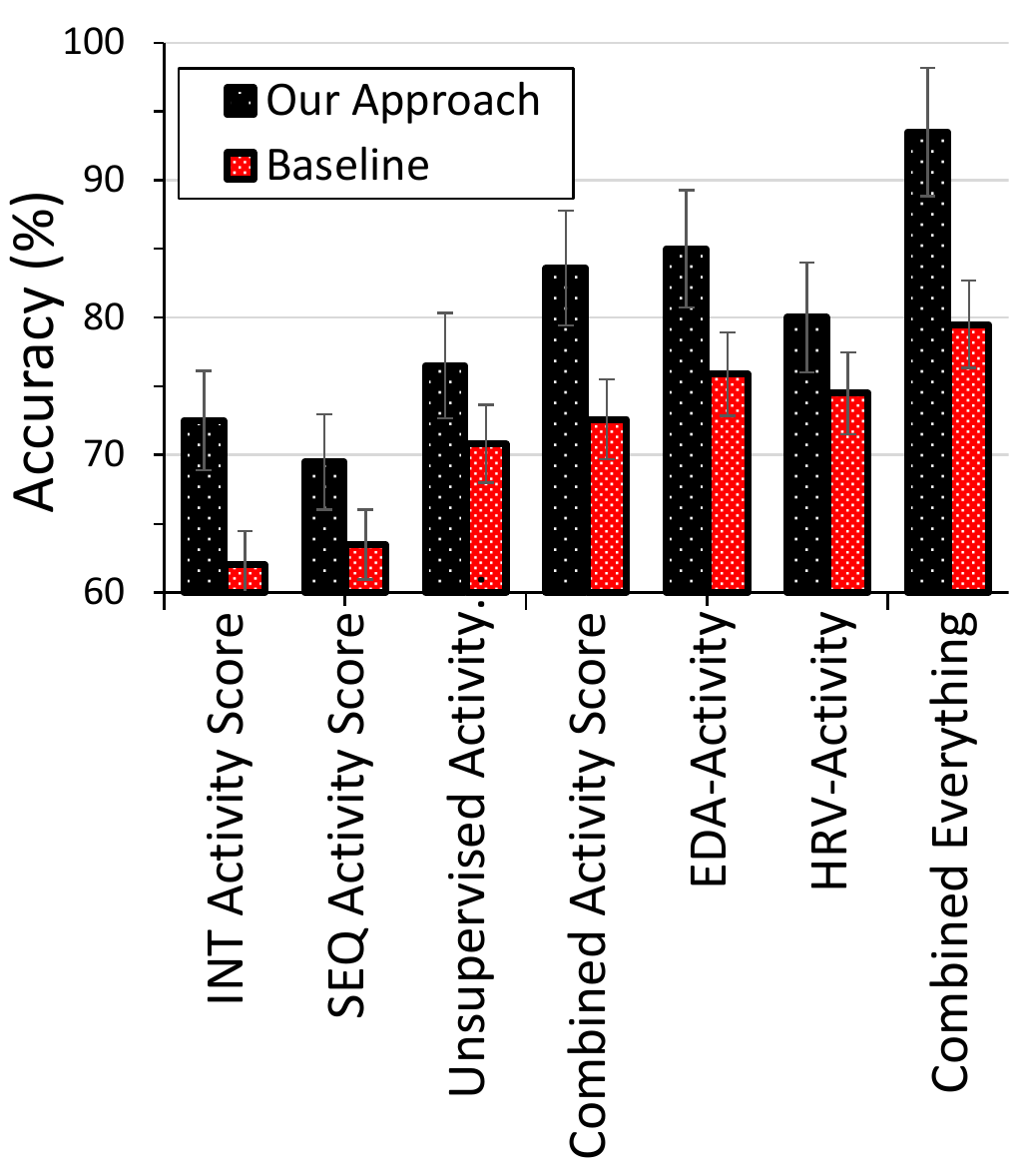,height=1.2in, width=1.7in}
   \vspace{-.15in}
\caption{Individual and combined classification accuracies comparison with baseline method for cognitive impairment status detection}
   \label{fig:combined_classification}
\end{center}
\end{minipage}
\begin{minipage}{0.23\textwidth}
\begin{center}
   \epsfig{file=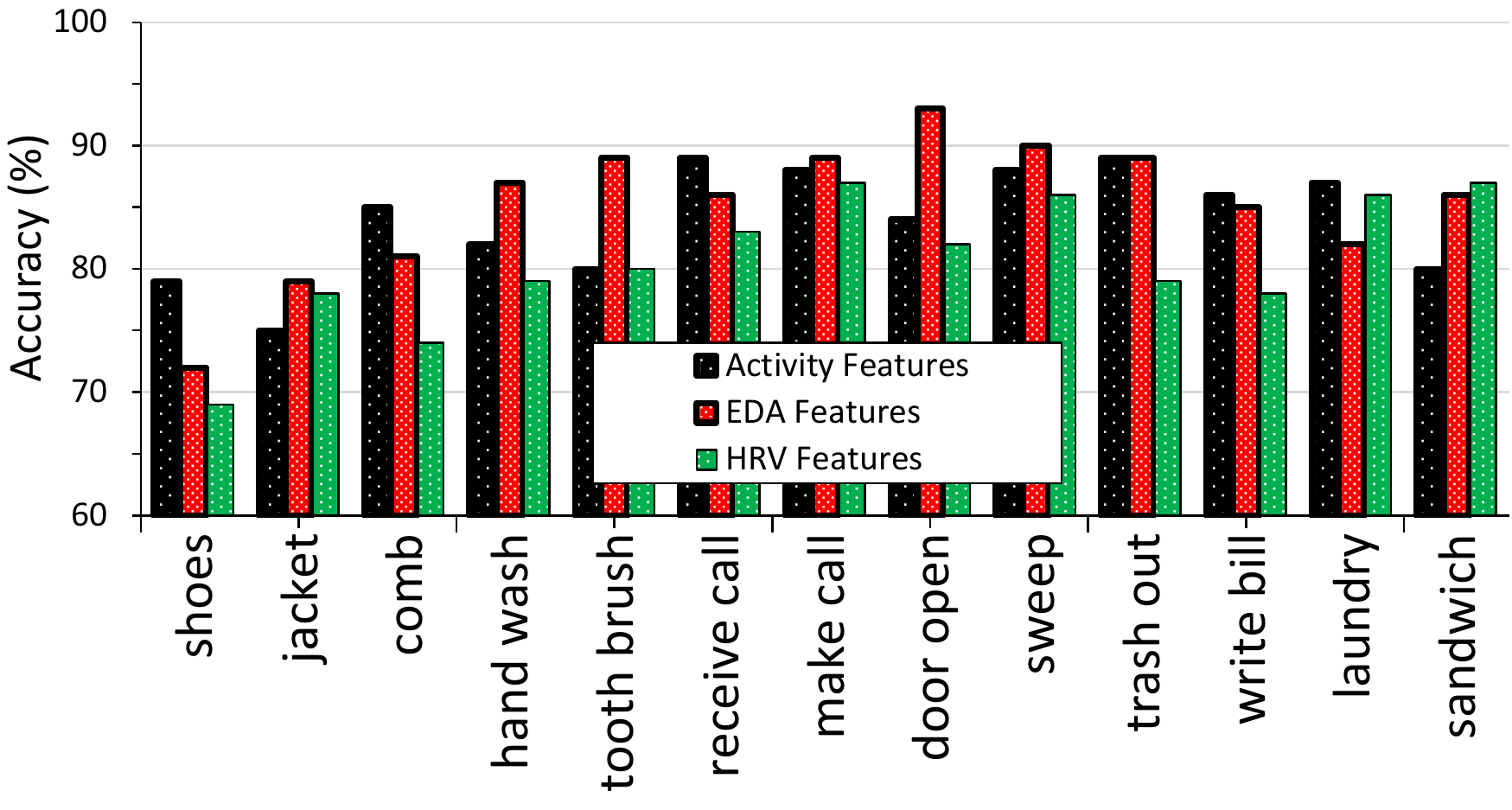,height=1.2in, width=1.7in}

\caption{Machine learning based cognitive health assessment accuracy for each complex activity in terms of activity, EDA and HRV features.}
   \label{fig:each_activity_cognitive_assessment}
\end{center}
\end{minipage}
\end{figure}

For combined classifier, we first applied sequential forward feature selection to find the best combinations of 1- 3 features for cognitive impairment classification group MCI, NCI and CI in terms of combined activity features (29 features), EDA-activity features (7 features) and HRV-activity features (8) features. Our final combined classifier (SMO based SVM algorithm \cite{cao06}) provides an accuracy of {\bf 93\%} in detecting the cognitive impairment status of older adults. Fig. \ref{fig:combined_classification} shows our proposed individual and combined methods outperform the baseline \cite{alam16} significantly (13\% improvement). Fig. \ref{fig:each_activity_cognitive_assessment} shows the cognitive impairment status prediction accuracy for each modality (activity feature, EDA and HRV) per individual complex activity.
\subsection{Discussion}
If we exclude the postural activities from automated activity performance scoring, we find reduced statistical correlation with original task completeness performance for \{NCI, MCI, CI\} participant group (INT 0.53*, SEQ 0.21' and unsupervised 0.49'). However, if we skip our proposed motion artifact removal stage, we find reduced statistical correlation with \{NCI, MCI\} and \{MCI, CI\} groups of participants (EDA and HRV correlations respectively \{0.51*, -0.51*\} and \{-0.53*,0.46\}). To test our proposed motion artifacts removal impact on EDA signals more rigorously, we choose 5 random participants, engage one expert motion artifact annotator to annotate motion artifacts segment on each participant's first 30 minutes of complex dataset using recorded video and apply both baseline and our methods to detect motion artifact segments. While baseline method achieves 75.5\% (FP rate 20.3\%) accuracy in detecting motion artifact segments, \emph{AutoCogniSys} outperforms achieving 89.9\% (FP rate 8.9\%) accuracy. In terms of experience, we have seen 100\% acceptance of wearing wrist-band,  71\% of acceptance for signing consent on using cameras and 0\% failure rate of collecting continuous data.
\section{Conclusion}
We propose, \emph{AutoCogniSys}, an IoT inspired design approach combining wearable and ambient sensors embedded smart home design, extensive signal processing, machine learning algorithms and statistical analytics to automate cognitive health assessment in terms of complex activity performances and physiological responses of daily events. Additionally, our postural activity detection approach in diverse population cum improved activity performance measurement and fundamental physiological sensor artifacts removal from physiological sensors help facilitate the automated cross-sectional cognitive health assessment of the older adults. Our efficient evaluation on each modality (physical, physiological and ambient) and each activity mode proves that any of the mode (say single activity and single sensor) also can provide significant improved cognitive health assessment measure.


\end{document}